\documentstyle[floats,multicol,aps,epsf]{revtex}
\input epsf
\input psfig

   \newcommand{\be}{\begin{equation}}
   \newcommand{\ee}{\end{equation}}
   \newcommand{\bea}{\begin{eqnarray}}
   \newcommand{\eea}{\end{eqnarray}}
   \newcommand{\upar}{\uparrow}
   \newcommand{\dn}{\downarrow}
   \newcommand{\delt}{\Delta_0}
   \newcommand{\ek}{\varepsilon_{k }}

   \newcommand{\om}{\omega_n}
   \newcommand{\bd}{\begin{displaymath}}
   \newcommand{\ed}{\end{displaymath}}

\begin{document}
\draft
\widetext
\twocolumn[\hsize\textwidth\columnwidth\hsize\csname @twocolumnfalse\endcsname

\title{Exact Density of States in a System of Disordered Two Dimensional Dirac Fermions in the Unitarity Limit: The d-wave Superconductor}
\author{Catherine P\'epin and Patrick A. Lee}
\address{Department of Physics, Massachussetts Institute of Technology, Cambridge, MA 02139}
\date{\today}

\maketitle \widetext
  \leftskip 54.8pt
  \rightskip 54.8pt
  \begin{abstract}
In this paper we present a method to compute the exact density of states induced by $N$ non magnetic impurities in a system of two dimensional Dirac fermions in the unitarity limit. We review the case of the $\pi$-flux phase of the Heisenberg model and also treat the disordered d-wave superconductor. In both case we find additional states in the gap with $\delta \rho (\omega) \simeq n_i/ \left | \omega \left ( \ln^2 |\omega/\Delta_0| + (\pi/2)^2 \right ) \right |$.
\par
  \end{abstract}
\vspace{0.2in}
]
\narrowtext

As a consequence of the breakdown of Anderson's theorem~\cite{anderson}, when impurity scattering violates the symmetry of the superconducting state, the superconducting energy gap is depleted and  impurtites act as strong pair breakers. This is the case in s-wave superconductors with magnetic impurities~\cite{abrikosov,kadanoff} for which there is creation of bound states in the gap. For the same reasons of symmetry violation, non magnetic impurities act as pair breakers in unconventional superconductors with higher orbital momentum such as d-wave superconductors. As a result, the measured low temperature properties of YBaCu$_2$0$_7$~\cite{hirsh1} displays a remarkable sensitivity to the presence of non magnetic impurities: the critical temperature, for example, is suppressed even in the lowest order in the disorder potential~\cite{alloul}.

\par
Furthermore the d-wave superconductor is special, due to the presence of gap nodes which prevents the complete freezing of scattering processes at low energy.

\par
The standard method to treat these problems of disorder~\cite{joynt7,rama} combines the T-matrix approximation with standard impurity averaging techniques. For three dimensional systems such as polar superconductors or heavy Fermion superconductors~\cite{joynt6} the standard perturbative approach is reliable. In the limit of low impurity concentration $n_i$, a perturbative expansion in $n_i$ leads to a finite density of states at the chemical potential~\cite{tsv2,joynt12}.

\par
For two dimensional systems (to which it is believed the high-T$_C$ cuprates belong) the standard procedure of averaging over impurities may be complicated by the appearance of logarithmic singularities in the perturbative expansion of the single electron self-energy~\cite{alexei}. Such a situation appears in a variety of two dimensional systems characterized by a Dirac-like canonical spectrum. In order to control the diagramatic expansion, all the diagrams with crossing impurity lines would have to be resummed which is technically impossible.

\par
In these systems the existence of a finite density of states induced by disorder is highly controversial~\cite{joynt}. 
Some non perturbative methods have been used to treat a weak disorder potential~\cite{alexei,mudry,rephettler}. The solution then depends on the symmetry of the pure system. In the special case of the d-wave superconductor the density of states still vanishes at low energy in the presence of disorder. 
On the other hand some perturbative self-consistent
calculations~\cite{hatsug,dhlee} and a non perturbative one~\cite{hettler} show a finite density of states.

\par
This issue of finite density of states is crucial for the conduction properties in the disordered compound. Indeed, if there exist states in the gap, possible anomalous overlaps between well separated impurities can induce a new conduction mechanism entirely through impurity wave functions in a so-called `impurity band'~\cite{balatsky,patrick}. This arises the question of localisation.

\par

In this paper, we treat the impurities in the unitary limit (infinite
 scattering potential $V_0$), but randomly distributed in space. In
this limit each impurity is identical, and create an extended bound
state at $E=0$. These states overlap, but a singular density of states
remains at zero energy.
Our approach is thus very different from models of random on-site energies
considered by others and really does not contradict them.

 We introduce a new method to calculate exactly the density of states induced by $N$ non magnetic impurities in a system of two dimensional Dirac fermions in the unitarity limit. 
We will outline the general features of the proof and present a different derivation than in our previous work~\cite{cath}.
The question of averaging will be discussed in details and we show how the result is exact provide the average over the different configurations of impurities is achieved at the end of the calculation.

\par
Our result can be applied to several systems of Dirac fermions in two dimensions. We review the case of the $\pi$-flux phase and explicitely treat the d-wave superconductor.

\par

The generic hamiltonian for a d-wave superconductor can be written 
\be
\label{eq115}
H= \sum_{ {\bf k } } \phi_{ {\bf k } }^\dagger \left[ \ek  \ \sigma_3 + \Delta_k \sigma_1 \right ] \phi_{ {\bf k } } \ ,
\ee
$$ \begin{array}{ll}
\mbox{with} & \ek= W \left( \cos{k_x} + \cos{k_y} \right) -\mu
\end{array} $$ 
$$ \begin{array}{ll}
\mbox{and} & \Delta_k = \Delta_0 \left( \cos{k_x} -\cos{k_y} \right)
\end{array} $$ 
which describes quasiparticles in the presence of the spin simglet order parameter $\Delta_k$. The $\sigma_i$ are the Pauli matrices in the particle-hole space and $\phi_{ {\bf k } }^\dagger = \left( c_{k,\upar}^\dagger, c_{-{ {\bf k}}, \dn} \right)$ is the spinor corresponding to the creation of a particle and a hole. We shall present our results in terms of a $d_{x^2-y^2}$ state eventhough our conclusions are general for any state where $\Delta_k$ vanishes linearly along a direction parallel to the Fermi surface.

\par

Instead of using the Nambu formalism, we work with the diagonalized
 version of~(\ref{eq115}) in order to access directly the properties
 of quasiparticles.

 The Bogoliubov transformation gives $c_{{\bf k} \upar}= u_{ {\bf k } }
\alpha_{ {\bf k } } - v_{ {\bf k } } \beta_{ {\bf k } }$ and
$c^\dagger_{-{ {\bf k}} \dn} = v_{ {\bf k}} \alpha_{ {\bf k}} +u_{
{\bf k}} \beta_{ {\bf k}}$ where $\alpha_{ {\bf k}}$ and $\beta_{ {\bf
k}}$ are the annihilation operators for quasiparticle and quasihole
respectively. The hamiltonian then reads $H= \sum_{{\bf k}} \omega_{ {\bf
k}} \left ( \alpha_{{\bf k}}^\dagger \alpha_{ {\bf k }} -\beta_{{\bf
k}}^\dagger \beta_{ {\bf k}} \right ) $ with $\omega_k =
\sqrt{ \ek^2 + \Delta_k^2}$. The coefficients $u_{ {\bf k}}$
and $v_{ {\bf k}}$ satisfy $u_{ {\bf k}}^2= 1/2 \left( 1+ \ek/
\omega_k \right)$, $v_{ {\bf k}}^2= 1/2 \left( 1- \ek/
\omega_k \right)$ and $u_{ {\bf k}} v_{ {\bf k}}= \Delta_k/(2 \omega_k)$. In order to simplify the calculation,
we consider that the spectrum is linearized around the four nodes
$\left( \pm {\tilde k}_F, \pm {\tilde k}_F \right)$ where close to 1/2
filling, ${\tilde k}_F= \left( \pi/2 - \mu \right)$ is close to
$\pi/2$.

\par
The disorder is introduced via a scalar scattering potential $V_0$
located randomly in the system $
H_I = V_0 \sum_{i=1}^N c^\dagger_i c_i
$, where $N$ is the number of impurities.
\par

 The lagrangian density can be written in the general form

\be
\label{eq1}
{\cal L}_{ {\bf k } {\bf k}^\prime}^{\alpha \beta}= G^{0 \ -1}_{ {\bf k }
\alpha} \delta_{\alpha \beta} \delta_{ {\bf k } {\bf k}^\prime} -V_0
\sum_{i=1}^N \left ( f^{\dagger i}_{ {\bf k } \alpha} f_{ {\bf k
}^\prime \beta}^i +  f^{\prime  \dagger i}_{ {\bf k } \alpha}
f^{\prime i}_{ {\bf k }^{\prime i}\beta}\right )  \ ,
\ee

This describes a system of two dimensional Dirac fermions in which $N$
non magnetic impurities located at $R_i$, $i=1$ to $N$, have been
introduced as repulsive scalar potentials of strenght $V_0$. Quasiparticles with positive energy ($\alpha=0$) and quasiholes with
negative energy ( $\alpha=1$)are described by the bare
Green's function $G^{0 \ -1}_{ {\bf k } \alpha}= i \om - (-1)^\alpha
\omega_k $. The factors $f_{ {\bf k } \alpha}^i$ and $f^{\prime i}_{ {\bf k } \alpha}$ come from the diagonalization of the initial hamiltonian. Two different terms usually arise from this diagonalization, leading to the sum in eq.~(\ref{eq1}).

\par
In the case of the d-wave superconductor, we have $f_{ {\bf k }
\alpha}^i = (-1)^\alpha t_{ {\bf k } \alpha} e^{ i {\bf k} \cdot {\bf
R}_i}$ and $f_{ {\bf k } \alpha}^{\prime i} =  t_{ {\bf k } \alpha} e^{ i {\bf k} \cdot {\bf R}_i}$ where $ t_{ {\bf k },0}= u_{ {\bf k}}$ and $t_{ {\bf k},1}= v_{ {\bf k}}$.

\par
In a previous paper~\cite{cath} we have studied the case of the
$\pi$-flux phase, considered as a model for the spin-gap phase of the
cuprates~\cite{autres1,khaliullin,cath}. Following the same
diagonalization as exposed in~\cite{cath}, we would get the factors:
$f_{ {\bf k} \alpha}^i =f^{\prime i}_{ {\bf k} \alpha} = e^{ i
{\bf k} \cdot {\bf R}_i}$ if $i$ belongs to sublattice A and $f_{ {\bf
k} \alpha}^i =f^{\prime i}_{ {\bf k} \alpha} = (-1)^\alpha e^{i {\bf k} \cdot {\bf R}_i} e^{ i \varphi_{ {\bf k}}}$ if $i$ belongs to sublattice B.

\par
Starting from the lagrangian~(\ref{eq1}), the equations of motion for the fully dressed Green's function can then be written
\be
\sum_{{\bf q} \gamma} {\cal L}_{ {\bf k} {\bf q}}^{\alpha \gamma}
G_{{\bf q} {\bf k}^\prime}^{\gamma \beta}=\delta_{ {\bf k} {\bf k}^\prime} \delta_{\alpha \beta} \ .
\ee

\par
We define ${\vec f}_{ {\bf k} \alpha}$ and ${\vec {\cal G}}_{ {\bf
k}^\prime \beta}$ as $N$ components vectors in the space of impurities
of components $f_{ {\bf k} \alpha}^i = u_{ {\bf k} \alpha}  e^{i {\bf
k} \cdot {\bf R}_i}$ and $ {\cal G}_{ {\bf k}^\prime \beta}^i =
\sum_{{\bf q} \gamma} f_{{\bf q} \gamma}^i G^{\gamma \beta}_{q {\bf k}^\prime}$ and respectively for ${\vec f}^\prime_{ {\bf k} \alpha}$ and ${\vec {\cal G}}_{ {\bf k}^\prime \beta}$.

\par
The Dyson equation (without any impurity averaging) can be written as:
\bea
G_{ {\bf k} {\bf k}^\prime}^{\alpha \beta} & = & G^0_{ {\bf k} \alpha} \delta_{ {\bf k} {\bf k}^\prime} \delta_{\alpha \beta}  \nonumber \\
& + & V_0 G^0_{ {\bf k} \alpha} \left ( {\vec f}^\dagger_{ {\bf k} \alpha} \cdot {\vec {\cal G}}_{ {\bf k}^\prime \beta} + {\vec f}^{\dagger \prime}_{ {\bf k} \alpha} \cdot {\vec {\cal G}}_{ {\bf k}^\prime \beta}^\prime \right )  \ , 
\eea
where the scalar product in the parentheses runs over the positions of impurities.

\par 
Solving with respect to ${\vec {\cal G}}$ and ${\vec {\cal G}}^\prime$ gives
\be
{\hat M}
\left ( \begin{array}{l}
  {\vec {\cal G}} \\ {\vec {\cal G}}^\prime \end{array} \right ) =
\left ( \begin{array}{l}
  G^0 {\vec f} \\  G^0 {\vec f}^\prime \end{array} \right ) \ ,
\ee 
$$\begin{array}{ll}
  \mbox{where} &   {\hat M} =\left ( \begin{array}{cc}
 {\hat I} + {\hat A} & {\hat B} \\
{\hat B} & {\hat I} + {\hat A} \end{array} \right ) \ ; \\ 
  & {\hat A}_{ij}= \sum_{ {\bf k} \alpha} G^0_{ {\bf k} \alpha}
  f^{\dagger i}_{ {\bf k} \alpha} f_{ {\bf k} \alpha}^i \\
\mbox{ and} & {\hat B}_{ij}= \sum_{ {\bf k} \alpha} G^0_{ {\bf k}
  \alpha} f^{\prime \dagger i}_{ {\bf k} \alpha} f^{\prime i}_{ {\bf k} \alpha}(R_j) \ .
\end{array} $$

\par
We thus get the T-matrix equation
\bea
G^{\alpha \beta}_{ {\bf k} {\bf k}^\prime} & = & G^0_{ {\bf k} \alpha} \delta_{ {\bf k} {\bf k}^\prime} \delta_{\alpha \beta} \\ 
& + & G^0_{ {\bf k} \alpha} 
\left ( \begin{array}{cc} {\vec f}_{ {\bf k} \alpha} & {\vec f}^\prime_{ {\bf k} \alpha} \end{array} \right ) {\hat T} \left ( \begin{array}{c} {\vec f}_{ {\bf k}^\prime \beta} \\ {\vec f}^\prime_{ {\bf k}^\prime \beta} \end{array} \right ) G^0_{ {\bf k}^\prime \beta} \ , \nonumber
\eea
with ${\hat T}= -V_0 {\hat M}^{-1}$.

\par
The additional density of states is given by $ \delta \rho = - 1 / \pi
Im \sum_{ {\bf k} \alpha} \delta G^{\alpha \alpha}_{ {\bf k k}}$. Using the equality $\partial {\hat G}^0/\partial i \om = {\hat G}^{0 \ 2}$ and the cyclicity of the trace we get $ \delta \rho (\omega ) = -1/ \pi Im Tr \left ( {\hat M}^{-1} \partial {\hat M} / \partial i \om \right ) $. The trace is over the positions of impurities.

\par
In our previous study~\cite{cath} we solved the problem by writing the term in the brakets as the derivative of a logarithm and evaluating the resulting determinant. Here we use another trick which consists in using the form of ${\hat M}^2$ to factorise the logarithmic divergence caracteristic of these problems. This derivation will make clear the question of averaging on disorder. We thus write
\be
\delta \rho (\omega) = \frac{-1}{2 \pi} Im Tr \left [ {\hat M}^{-2} \partial {\hat M}^{ 2} / \partial i \om \right ]
\ee
and carefully study the form of ${\hat M}^2$.

\par
The solution of the problem depends on the actual form of $M_{ij}$.

\par
 
As for the $\pi$-flux phase, we can show that $M_{ij} \approx 1/
R_{ij}$. This fact was previously acknowledged by other
othors~\cite{balatsky,repbalat} who found an overall factor of the
order of $( \Delta_0 / W ) 1/R_{ij}$.

Let's first consider the case of symmetric bandwidths for positive and
negative energies. This means we are at half filling for the
superconductor ($\mu=0$).
We show below how this $1/R$ value arises in the case where $ W = \Delta_0$.

We get
\[
A_{ij}= \sum_{ {\bf k} \in BZ} \frac{ u_{ {\bf k}}^2 e^{ i {\bf k } \cdot { \bf
R}_{ij} } } {  i \om  - \omega_k } + \sum_{ {\bf k} \in BZ} \frac{ v_{ {\bf k}}^2 e^{
i {\bf k } \cdot { \bf R}_{ij} } } {  i \om  + \omega_k }
\]
so that
\bea
A_{ij} & = & \sum_{ {\bf k} \in BZ} \frac{ i \om e^{ i {\bf k } \cdot
{ \bf R}_{ij} } } {  i \om  - \omega_k } + \sum_{ {\bf k} \in BZ} \frac{
\ek e^{ i {\bf k } \cdot { \bf R}_{ij} } } {  i \om +  \omega_k } \nonumber \\
 & = & 2 \pi {\cal F}^0 (R_{ij} ) \frac{ \omega}{\Delta_0^2} \ln \left | \frac{ \omega}{\Delta_0} \right | + \frac{ 2 i \pi }{ R_{ij} } {\cal F}^1 (R_{ij}) \ ,
\eea
with $ {\cal F}^0 (R) =2 \left [ \cos \frac{\pi }{2} ( R_x +R_y ) + \cos \frac{\pi}{2} ( R_x - R_y) \right ]$ and 
\bea
{\cal F}^1 (R)&  = & 2 \left [ \sin \frac{\pi }{2} ( R_x +R_y ) \left ( \cos \varphi + \sin \varphi \right ) \right . \nonumber \\
  &  + & \left . \sin  \frac{\pi}{2} ( R_x - R_y) \left ( \cos \varphi - \sin \varphi \right ) \right ] \nonumber \ ,
\eea
 where $\varphi$ is the angle between ${\bf R}$ and the ${\vec x}$ axis.

\par 
For ${\hat B}$ we write
\[
B_{ij}= \sum_{ {\bf k} \in BZ} \frac{ u_{ {\bf k}} v_{ {\bf k}} e^{ i {\bf k } \cdot { \bf
R}_{ij} } } {  i \om - \omega_k } + \sum_{ {\bf k} \in BZ} \frac{ u_{ {\bf k}} v_{ {\bf k}} e^{
i {\bf k } \cdot { \bf R}_{ij} } } { i \om + \omega_k }
\]

so that we find
\be
B_{ij} = \frac{ 2 i \pi}{ R_{ij}} {\cal F}^2 (R_{ij} ) \ ,
\ee
with 
\bea
{\cal F}^2 (R) & = &2 \left [ \sin \frac{\pi }{2} ( R_x +R_y ) \left ( \cos \varphi - \sin \varphi \right )  \right . \nonumber \\
 & + & \left . \sin  \frac{\pi}{2} ( R_x - R_y) \left ( \cos \varphi + \sin \varphi \right ) \right ] \nonumber \ .
\eea

\par
The key point in the evaluation of the density of states is that
${\hat M}^2$ shows a logarithmic divergence.

 We factorize explicitely this divergence by taking

\be
{\hat M}^2 = \ln \left | \frac{\delt}{\om} \right | \ {\hat S}
\ee
where ${\hat S}$ is a matrix which depends on the particular configuration of impurities we work with. Two terms emerge in the density of states:

\be
\delta \rho (\omega) = \frac{-1}{\pi} Im Tr \left [ \ln^{-1} \left |
\frac{\om}{\delt} \right | \frac{\partial \ln | \om/ \delt |}{\partial
\omega} \ {\hat I} + {\hat S}^{-1} \frac{\partial {\hat S} }{\partial
i \om } \right ] \ .
\ee

Note that the first term doesn't depend on the particular configuration of impurities. The average over disorder will act only on the second term. After analytic continuation, we find per unit of volume

\be
\label{eq11}
\delta \rho (\omega ) = \frac{n_i}{ \omega \left ( \ln^2 \left | \omega / \delt \right | + (\pi/2)^2 \right ) } - \frac{2}{\pi V} Im \left \langle Tr {\hat S}^{-1} \frac{\partial {\hat S}}{\partial \omega} \right \rangle \ ,
\ee
where the brakets denote the average over disorder.

\par
We argue now that the contribution coming from ${\hat S}$ is
negligible. For this purpose, we make an overestimate of ${\hat S}$ with the $N/2 \times N/2$ matrix ${\hat S}^a$ satisfying
$S^a_{ij} =1$ if $R_{ij} < \delt /|\om|$ and $0$ elsewhere. In
this approximation we have overevaluated the long distance intercations
between impurities by replacing the slow decay of $S_{ij}$ which
vanishes at $R_{ij}= \left |  \delt / i \om \right |$ by a step
function with a cut-off at$R_{ij}= \left | D/ i \om \right |$. This
special form of ${\hat S}$ thus leads to an upper bound for the
averaged second term in~(\ref{eq11} ). Our aim is now to prove that it
is negligible as compared to the first term. $\partial {\hat S}^a /
\partial \omega$ is then the matrix which elements $\partial S^a /
\partial \omega |_{ij}$ vanish everywhere except on the external
boundary of the circle of radius $R_{ij} = \left | \delt / \om \right |$.

\par
The main difficulty in the inversion of ${\hat S}^a$ is that two
circles of radius $\delt /|\om|$ centered around two points $i$ and
$j$ very close to each other will overlap, leading to the same number
of non zero coefficients in the lines $i$ and $j$ of ${\hat S}^a$. In
order to differentiate the sums $\sum_{ {\bf k}} S^a_{i {\bf k}} S_{
{\bf k}i}^{a \ -1}$ and $\sum_{ {\bf k}} S^a_{i {\bf k}} S_{ {\bf k}j}^{a \ -1}$ we have thus used the external boundary of the circle to compensate its volume in ${\hat S}^{a \ -1}$.

We take ${\hat S}^{a \ -1}_{ij} =\left  (|\om|/( \sqrt{\pi} \delt ) \right
) \ U_{ij}$. For all $j$ inside a circle of radius $\delt /|\om|$
around the point $i$, $U_{ij}$ is a random configuration of $\pm 1$ so
that $\sum_j U_{ij} \sim \sqrt{\pi} \delt /|\om|$. In addition all the points $j$ situated in the external boundary of this circle have $U_{ij}= -1/(2 \sqrt{\pi})$. Elsewhere $U_{ij}=0$.

\par
We thus get ${\hat S}^{a \ -1} \left ( \partial {\hat S}^a / \partial \omega \right ) = \sqrt{\pi}$ wich is negligible as compared to the first term. Thus this term is of second order as compared to the first term in~(\ref{eq11}).

\par
Provided ${\hat M}^2$ displays a logarithmic divergence, we can factorize this divergence and show that the exact density of state in the unitarity limit and in the limit of low frequencies is
\be
\delta \rho (\omega ) \simeq \frac{n_i}{ |\omega| \left ( \ln^2
|\omega/ \delt | +(\pi/2)^2 \right )} \ .
\ee

In the case where $\Delta_0 \neq W $ the logarithmic singularity gets
an overall prefactor $\Delta_0/ W $~\cite{balatsky} which leads to the most general form at low energy
\be
\label{eq22}
\delta \rho ( \omega ) \simeq \frac{\Delta_0}{W} \frac{
n_i}{| \omega | \left ( \ln^2 \left | \omega / \Delta_0  \right | + (\pi/2)^2 \right ) } \ .
\ee

The equation~(\ref{eq22}) is thus the most general expression for the
density of states induced by non-magnetic impurities in the unitarity
limit, and for energy bands symmetric for positive and negative
energies.

\par
Note that this expression is normalisable. If the impurities were totally uncorrelated, we would find $N$ $\delta$-functions at $\omega=0$. Here we see clearly the overlap of impurity states which leads to a broadening of the $\delta$ functions. The interacting states repeal each other, but not enough to kill the divergence. This level repulsion was found by other methods~\cite{alexei} for Dirac fermions in two dimensions in presence of disorder, but away from unitarity limit.

\par
What happens if the bandwiths are not symmetric anymore, that is if
$\mu \neq 0$ but still $\mu \ll \Delta_0$ ?

First the nodes are moved away from the point $ \left ( \pm \pi/2, \pm
\pi/2 \right )$ so that transversal nodes are now separated y the
vectors $ {\bf Q}= \left ( \pi (1-\delta), \pi (1-\delta) \right )$
and ${\bf Q}^*= \left (- \pi (1-\delta), \pi (1-\delta) \right )$
where $ \delta= \mu / \Delta_0 $ and $\mu$ is the increase in the chemical
potential. This leads to a change of the phase factors in $A_{ij}$ and
$B_{ij}$. Namely we get $ {\cal F}^0 (R) =2 \left [ \cos \left ( {\bf Q} \cdot
{\bf R} /2 \right ) + \cos \left ( {\bf Q}^* \cdot {\bf R} /2 \right ) \right ]$; 
\bea
{\cal F}^1 (R)&  = & 2 \left [ \sin \left ( {\bf Q} \cdot {\bf R} / 2
\right ) \left ( \cos \varphi + \sin \varphi \right ) \right . \nonumber \\
  &  - & \left . \sin \left ( {\bf Q}^* \cdot {\bf R} / 2 \right ) \left ( \cos \varphi - \sin \varphi \right ) \right ] \nonumber \ ,
\eea and

\bea
{\cal F}^2 (R) & = &2 \left [ \sin \left ( {\bf Q} \cdot {\bf R} / 2
\right ) \left ( \cos \varphi - \sin \varphi \right )  \right . \nonumber \\
 & - & \left . \sin \left ( {\bf Q}^* \cdot {\bf R} / 2 \right ) \left ( \cos \varphi + \sin \varphi \right ) \right ] \nonumber \ .
\eea

As $\delta$ is a small parameter, this change in the phase won't affect the existence of the logarithmic divergence in ${\hat M}^2$.

\par
More importantly, away from half filling, the bands of quasi particles and quasi holes become asymmetric to account for the removing of particles in the system. This gives additional terms in $A_{ij}$ and $B_{ij}$ of the form $\sum_{1 < |k| < 1+\delta} e^{i {\bf k} \cdot {\bf R}} / ( i \om - \omega_k)$. We show that~\cite{note} their contribution  in ${\hat M}^2_{ij}$ is of order of $ \delta  J_0(R_{ij})/ \Delta_0^2$ where $J_0$ is the Bessel function of rank zero.  This doesn't affect the logarithmic divergence at low energies.

\par
The density of states is thus stable with respect to a small perturbation of the chemical potential.

\par
The result~(\ref{eq22}) holds for any
system of Dirac fermions in two dimensions, provided the hopping
matrix between inpurities goes like $1/ R_{ij}$. 
Thus it seems to be generic for two dimensional Dirac fermions in presence of Poissonian disorder. Indeed the same result as~(\ref{eq22}) has been obtained for a system of electrons in a strong magnetic field in two dimensions, when the disorder is Poissonian~\cite{brezin}.

\par
This result has to be compared with one dimensional spin-Peierls and two-leg ladder compounds for which non magnetic impurities create a Dyson-like singularitty $\delta \rho (\omega) \sim 1/ \left ( | \omega | \ln^3 | \omega / D | \right )$~\cite{gogolin,dyson}. In one dimension, the Dyson singularity is a signature of delocalised states in the system: there is a direct relation between the localisation lenght and the density of states~\cite{thouless} which shows the divergence of the localisation lenght at low energies. In two dimensions this theorem doesn't hold anymore and the question of localisation is still open.

We would like to thank A. M. Tsvelik, R. Joynt, M. H. Hettler,
A.V. Balatsky, P. Coleman and J. Chalker for useful discussions
related to this work.

 This work is supported by NSF Grant No. DMR-9523361 and by (CP) a Bourse Lavoisier.

	\vspace{-.5  cm} 



\end{document}